\documentclass[aps,prb,amsfonts,showpacs,floats,floatfix,twocolumn,nofootinbib,byrevtex]{revtex4}
\usepackage{epsfig}
\usepackage{amsmath}
\usepackage{amsfonts}

\usepackage{color}
\usepackage{graphicx,slashbox}

\usepackage{ulem}

\begin{document}

\title{Production of intense beams of mass-selected water cluster ions and theoretical study of atom-water interactions}
\author{Z.P. Wang$^{1,2,6}$, P.M. Dinh$^{1,2}$,P.-G. Reinhard$^3$, E. Suraud$^{1,2}$}
\affiliation{
  1) Universit\'e de Toulouse; UPS; Laboratoire de Physique Th\'eorique
  (IRSAMC); F-31062 Toulouse, France
\\
  2) CNRS; LPT (IRSAMC); F-31062 Toulouse, France
\\
  3) Institut f\"ur Theoretische Physik,
  Universit{\"a}t Erlangen, D-91058 Erlangen, Germany,\\
  6) The Key Laboratory of Beam Technology and Material
Modification of Ministry of Education, College of Nuclear Science
and Technology, Beijing Normal University, Beijing 100875,
People's Republic of China}

\author{G. Bruny$^4$, C. Montano$^4$, S. Feil$^4$, S. Eden$^{4,\footnote{   Dept. Physics and Astronomy, The Open University (OU), Walton
hall, Milton Keynes, MK76AA, UK
}}$,
H. Abdoul-Carime$^4$, B. Farizon$^4$, M. Farizon$^4$, S. Ouaskit$^{4,\footnote{Universit\'e Hassan II-Mohammedia, Faculté des Sciences Ben M'Sik (LPMC),
B.P.7955 Ben M'Sik, Casablanca, Morocco
}}$, T.D. M\"ark$^5$}
\affiliation{
  4) Universit\'e de Lyon, F-69003, Lyon, France; Universit\'e Lyon 1, Villeurbanne;
\\
CNRS/IN2P3, UMR5822, Institut de Physique Nucl\'eaire de Lyon; F-69622
Villeurbanne
\\
  5) Institut f\"ur Ionenphysik und Angewandte Physik,
Leopold Franzens Universit\"at,
Technikerstrasse 25, A-6020 Innsbruck, Austria}
\date{\today}
\begin{abstract}
The influences of water molecules surrounding biological molecules during irradiation with heavy particles (atoms,ions) are currently a major subject in radiation science on a molecular level. In order to elucidate the underlying complex reaction mechanisms we have initiated a joint experimental and theoretical investigation with the aim to make
direct comparisons between experimental and theoretical results.
As a first step, studies of collisions of a water molecule with
a neutral projectile (C atom) at high velocities ($\geq 0.1 $a.u.), and with a charged projectile (proton) at low velocities ($\leq 0.1 $a.u.) have been studied within the microscopic framework. In particular, time-dependent density functional theory (TDDFT) was applied to the
valence electrons and coupled non-adiabatically to Molecular dynamics
(MD) for ionic cores. Complementary experimental
developments have been carried out to study projectile interactions with accelerated ($\leq10$keV) and mass-selected cluster ions. The first size
distributions of protonated water cluster ions H$^+$(H$_2$O)$_{\rm n}$(n=2-39) produced using this new apparatus are presented.

\end{abstract}

\pacs{xxx}

\maketitle

\section{Introduction}
\label{sec:goal} While the use of ionizing radiation is
well-established, notably in therapies and analytical techniques, each
new development opens a field of investigation around the possible
dangers to our health and the environment \cite{Paretz,Jacob}. Concurrently, besides applications and risk evaluations, the effect of ionizing
radiation in biomolecular nanosystems is emerging as a major area of
research, both on a fundamental level and as a source for experimental
and technical innovations. The irradiation of biomolecular nanosystems
in the gas phase under single collision conditions represents a significant new subject in radiation
science. From an experimental standpoint, it has great potential for
the development of new analytical techniques and synthesis
techniques. Complementary theoretical progress can provide new
descriptions of radiative energy transfer mechanisms on the molecular
scale, opening new perspectives for the elucidation of the radiation
dose in living systems.

Our principle goal is to investigate microscopic mechanisms responsible for irradiation effects in biological molecules with a particular focus on the influence of water molecules in the environment of these molecules.This is achieved here in a joined experimental and theoretical effort
with the aim to make direct comparisons between experimental and
theoretical results. The originality of this work lies in the
ability to quantify the role of a biomolecule's immediate environment
in proton radiation-induced processes through the precise control
of the number of associated water molecules. This is a key aspect from
the fundamental physics point of view and it is also crucial to
allow realistic and detailed comparisons between experiments and theory. The path towards such an ambitious
goal is long and due to the many elementary processes involved one
needs to validate the various experimental and
theoretical aspects. First of all, we consider here the
 simplest case of a single water molecule interacting with a
projectile, which constitutes a prerequisite for any further
developments. The next step involving an assemble of water molecules around  a biomolecule raises no major theoretical difficulty because the number of molecules can be efficiently controlled. Monitoring this number experimentally constitutes a key issue for further experimental
developments and for comparisons with theory.

In previous experiments at the Institut de Physique Nucl\'eaire de
Lyon, the ionization of gas-phase water molecules by
protons\cite{Gob01} and neutral hydrogen\cite{Gob06} impact has
been studied in the velocity range of 0.9-2.5 a.u.(units of the Bohr velocity), coinciding
with the Bragg peak maximum for energy deposition by an ion beam
in an absorbing medium. Mass-analyzed H$_2$O$^+$ and fragment ions
of the water molecules were detected in coincidence with the
projectile following ionization in single collision conditions.
The determination of the charge state of the projectile after the
collision enabled ionization processes to be separated on the
basis of charge transfer between the target molecule and the
projectile. Thus it was possible to measure branching ratios and
absolute cross sections for water ionization with and without the
capture of an electron by the incident proton.

The present experimental challenge is to study interactions between water molecules in a cluster following proton impact induced excitation/ionization
(20-150 keV)of water prior to dissociation. Accordingly a new experimental system has been
developed to investigate the proton irradiation of mass-selected
protonated water clusters. Many different techniques have been
used successfully to generate ensembles of water cluster ions
including electrospray ionization \cite{Led97,Tom01}, electron
impact ionization \cite{Ech84}, corona discharge ionization
\cite{Ech84}, chemical ionization \cite{Her82} and electrospray
droplet impact \cite{Hua06,Mor06}. The novelty of the present
experimental system lies in the acceleration (up to 10 kV) and
mass-selection of  molecular and cluster ions of water  prior to
0.9-2.5 a.u. collisions with protons.

The microscopic description of radiation induced processes requires an
explicit dynamical account of electronic degrees of freedom which
respond first in the present collisions. Moreover, it is
necessary to treat electrons in a non-adiabatic way and to allow
for ionization and/or electron transport. This invalidates most
calculations based on the Born-Oppenheimer (BO) approximation
except in some specific situations. Indeed, depending on the
characteristics of the ionizing projectile (charge, velocity), one
can treat the problem in a simplified manner by decoupling
electronic and ionic dynamics. A typical example is the case of
high velocity charged projectiles in which ions can be safely
considered as frozen and the dynamics reduced to the electronic
response, at least for short times. Another example is the case of low
velocity neutral projectiles for which a ground state
(Born-Oppenheimer, BO) treatment is acceptable. To the best of our
knowledge, low velocity charged and high velocity neutral
projectiles can nevertheless not be treated by the usually available
calculations \cite{Koh08}, and certainly not in the framework of a
unique theoretical approach \cite{Gai07}. Fortunately, in the case of cluster
dynamics, several approaches \cite{Cal95a,Yab96,Cal00,Dom98b,Fen04}
have been developed which consider this question of
coupled electronic and ionic dynamics in relation to irradiation 
by intense laser fields. 
A coupling of the optically active spot to a large environment
can be added in a hierarchical approach \cite{Feh06a,Bae07a}.
In the present work we have adapted the
non adiabatic approach of Calvayrac et al. \cite{Cal00} to the
case of an organic molecule and applied it to realistic irradiation
scenarios. The method of Calvayrac et al. \cite{Cal00} contains as
limiting cases pure electron dynamics and BO dynamics
(Car-Parinello dynamics) and can thus describe collisions with
high-velocity charged or low-velocity neutral projectiles.
Moreover, at variance with currently available approaches, it
enables complementing cases to be described  such as low-velocity
charged and high-velocity neutral projectiles. Indeed such a non
adiabatic approach places no restriction on the velocity
or charge state
of the projectile and thus offers an
unified picture of many possible irradiation scenarios.

\section{Experimental set-up}
\label{sec:exp}

\begin{figure}[htbp]
\begin{center}
\includegraphics[width=\linewidth]{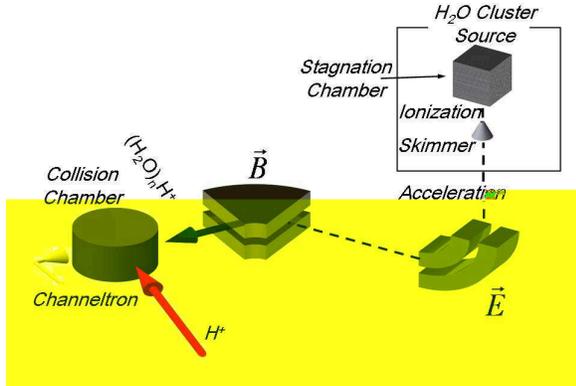}
\caption{Schematic view of the experiment}
\label{fig:exp_setup}
\end{center}
\end{figure}

The present experimental system is shown schematically in Figure
\ref{fig:exp_setup}. All measurements are taken with a
two-sector-field mass spectrometer of EB geometry combined with an
electron-impact ion source. Neutral clusters are produced by
expanding water vapor from a stagnation chamber at a temperature
of around 100°C and a pressure of approximately 1 bar through a
pin-hole nozzle (20 $\mu m$ diameter) into vacuum. In this
isentropic expansion the temperature drops rapidly with increasing
distance from the nozzle leading to super-saturation of the water
vapor and subsequent clustering. The clusters are ionized by
electrons of 50 eV (experimentally available range from about 0-50 eV) and the pressure in this
region is maintained to {2x10$^{-4}$} mbar by a 1200l/s
oil diffusion pump. The resulting positive ions and cluster ions
(charge state q) are immediately extracted from the
center of the ion source by a weak penetrating electric
field. The extracted ions pass through a skimmer and are
accelerated to (6xq) kV. After the ionization chamber
the ions pass a skimmer. An Einzel lens and four pairs of
deflection plates are used to direct the beam to the focal point
of a hemispherical electric sector field with maximum
transmission. The electric sector field serves to select ions
according to their kinetic energy, thus increasing the energy
resolution of the beam. The KE-resolved ion beam then passes
through a magnetic sector field for ion selection according to
momentum. Finally, the mass-selected water cluster ions
H$^+$(H$_2$O)$_{\rm n}$ are focused into the {\it collision chamber} and
detected through a 1 mm collimator by a channeltron electron
multiplier operated in a counting mode. Further experimental developments are currently being carried out in order to intersect the cluster ion beam 
with an intense beam of 0.9-2.5
a.u. protons or a jet of atomic or molecular gas (He, N$_2$...).

\section{Theoretical methods}
\label{sec:theo}

In order to perform microscopic simulations of dynamical
processes, we employ time-dependent density functional theory
(TDDFT) for the electrons combined with classical molecular
dynamics (MD) for the ionic cores. The TDDFT is used at the level
of the local-density approximation (LDA) together with an
average-density self-interaction correction (ADSIC) to achieve the
correct ionization potentials (IP), a feature which is crucial to
describe electron emission correctly. Wavefunctions and fields are
represented on a 3D coordinate-space grid of dimensions 
72*72*64. Absorbing boundary conditions are used to remove
outgoing electrons. Thus the total number of electrons  $N=N(t)$
decreases in time. The number of escaped electrons
$N_\mathrm{esc}=N(t\!=\!0)-N(t)$ is a measure of average
ionization. The details of this method are presented elsewhere
(e.g., \cite{Cal00,Rei03a} and for ADSIC \cite{Leg02}).

The coupling between electrons and ionic cores is achieved by
Goedecker-type\cite{Goe96a} non-local pseudo-potentials (PsP). The
original parameterizations employ different Gaussian widths for
each material and contribution. This hampers numerical precision
in coordinate space grids. We have refitted new PsP parameters
for the smallest width the same value of 0.412
a$_0$ in all terms (local and non-local) and for all elements
involved in the study. Further details will be given in a
forthcoming publication.

\section{First Results}
\label{sec:res}

\subsection{ Experimental production of size resolved protonated water clusters}
\label{sec:res_exp}

\begin{figure}[htbp]
\begin{center}
\includegraphics[scale=0.42, bb= 0 280 900 620]{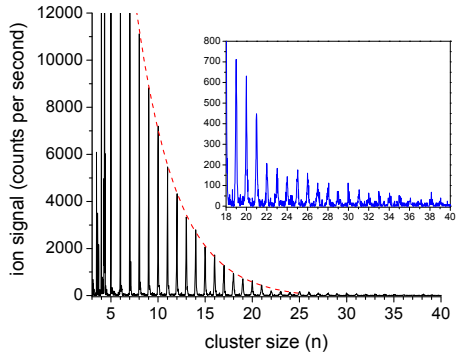}
\caption{Spectrum of mass-selected  water cluster ions at 6 keV energy
obtained by scanning the magnetic sector field (see figure \ref{fig:exp_setup}).}
\label{fig:water_spectrum}
\end{center}
\end{figure}
Figure 2 shows a typical mass spectrum obtained for the water cluster beam by scanning the magnetic sector field.
This spectrum was recorded for an acceleration voltage of 6 kV and
for a water vapor pressure and temperature of 1.1 bar and 100 C in
the stagnation chamber, respectively. The water monomer ion peak is not
shown in the spectrum as its intensity is too high leading to
saturation of the channeltron detector. The observed series of
peaks spaced by 18 amu / unit charge is attributed to protonated
water clusters comprising n molecules (n = 2-39). Water cluster
ion intensities decrease exponentially (red dashed line in Fig. 2)
as the number of water molecules increases. The other peaks
correspond to components arising from the residual gas in the
cluster source. The insert in figure 2 shows the higher-mass peaks
(n=18-39) in greater detail. These peaks correspond to
H$^+$(H$_2$O)$_{\rm n}$ (n = 18-39) cluster ions. The present size
distribution of cluster ions is rather similar to that observed
by Castleman and co-workers \cite{Her82}. The present results are plotted in Figure 3 also on a logarithmic scale. In this case a clear drop in intensity can be seen for ion production above cluster size n=21. This is consistent with previous results \cite{Her82} supporting a {\it magic number} (a particularly stable configuration) at this cluster size. 
\begin{figure}[htbp]
\begin{center}
\includegraphics[scale=0.38, bb=0 270 540 670]{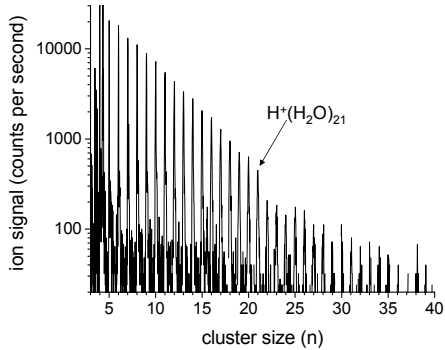}
\caption{Spectrum of mass-selected  water cluster ions at 6 keV energy
in a log scale}
\label{fig:water_spectrum}
\end{center}
\end{figure}   It is worth noting that the intensity of the
mass-analysed cluster ion beams achieved with the present set-up is rather high. For instance, the intensity of
the protonated water trimer beam measured through the 1
mm-collimated channeltron is 5000 counts/s. In a next step these mass selected cluster ion beams will be crossed by an intense proton beam and
the resulting collision events will be analyzed using a molecular
imaging detection system.

\subsection{Theoretical studies of the irradiation of water molecules}
\label{sec:res_theo}

As discussed in section \ref{sec:goal} the ultimate aim of our
combined theoretical and experimental project
is a systematic investigation of collision induced processes in H$^+$(H$_2$O)$_{\rm n}$
cluster cations. To achieve this ambitious
goal, it is necessary to validate the employed methods
at intermediate steps. As the key building block of
H$^+$(H$_2$O)$_{\rm n}$ clusters,
 the neutral H$_2$O molecule provides the starting point for our theoretical investigation.
Systematic studies of small H$^+$(H$_2$O)$_{\rm n}$ (n = 1-3) will
be presented elsewhere. The first models developed here will describe the collision of an atom or ion with an H$_2$O molecule. As
observables, we study the ionization and the motion of the various
atoms in the system. In particular we have explored the two
limiting cases of low-velocity neutral and high-velocity charged
projectiles for which one can reduce the dynamics to
Born-Oppenheimer or pure electronic motion, respectively. We have
checked that our non-adiabatic approach enables us to recover
these limiting cases, as expected. In the present studies, we find
that non-adiabatic effects do indeed remain negligible which a
{\it posteriori} validates previous investigations. We thus focus
on the cases of low-velocity charged and high-velocity neutral
projectiles for which an accurate coupling between ionic
and electronic dynamics becomes compulsory. We illustrate the
capabilities of our approach on such cases. 

\subsubsection{Neutral atom impact upon H$_2$O at high-velocity }
\label{sec:res_theo_neutral}
\begin{figure*}[htbp]
\begin{center}
\includegraphics[width=0.85\linewidth]{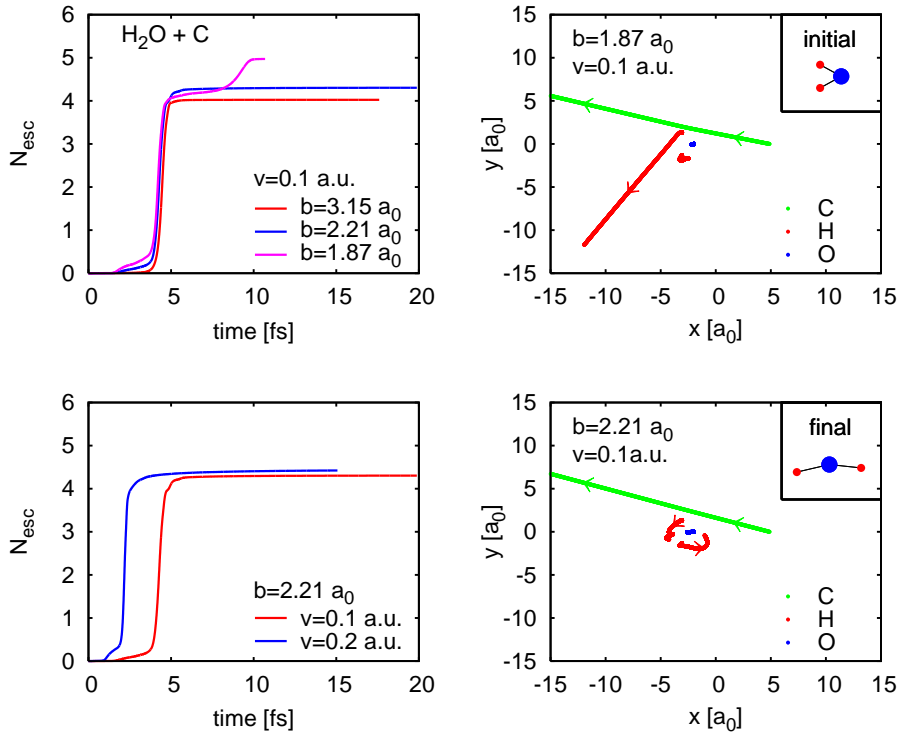}
\caption{Modeled collision between a water molecule and a C atom
at various high velocities and impact parameters $b$. Left
panels~: number of escaped electrons as a function of time. Right
panels: ionic trajectories in the $x-y$ plane; the arrows indicate
the time evolution; the insert in the top right panel shows the
initial configuration of H$_2$O for all calculations, while that
in the bottom right panel presents the final configuration only
for the case $v=0.1$~a.u. and $b=2.21$~a$_0$.} \label{fig:fastC}
\end{center}
\end{figure*}
Fig.~\ref{fig:fastC} shows results for a collision of a fast
(v=0.1-0.2 a.u.) neutral C atom with an H$_2$O
molecule.
 
The water molecule lies
in the scattering plane. The upper left panel shows the time
evolution of ionization for impact velocity
$v$=0.1$\,\mathrm{a.u.}$=20$\,\mathrm{a}_0$/fs and varying impact
parameter $b$. The sudden jump to $N_\mathrm{esc}\approx4$ is
caused by the C atom leaving the numerical box and carrying its
four active electrons away. The net ionization of the water
molecule is the difference $N_\mathrm{esc}$-4. There is
negligible ionization (and excitation) of the water molecule for
the largest $b$ (3.15 a$_0$). For $b=2.21$ a$_0$, however, we see a
loss of about 0.2 charge units whereas for the smallest b
(1.87 a$_0$) one full electron is lost. The reason for this
becomes clear when we consider the right upper panel showing the
trajectories in the scattering plane for the smallest impact
parameter. The C atom travels from right to left through the box
with negligible deviations from a straight line, first triggering
ionization by about 0.2 charge units as shown in the left-hand
panel. At the point of closest impact, the C atom hits one of the
H atoms and transfers a sufficiently large momentum that
the H atom dissociates from the molecule and leaves the
box taking the attached 0.8 electrons away.
\\
\\
The lower right panel shows the trajectories for the impact
parameter. There is again a strong impact on the H atom closest to
the projectile. As for the smaller impact parameter, there is
still a significant ionization of about 0.2 charge units. However,
the transferred momentum is sufficiently low that the atom remains
bound in the molecule. The H atom rotates around the O
atom and pushes the other H atom into the same rotation such that
the structure of molecule is essentially conserved. Thus
atom impact has triggered here just a strong rotational motion.
Finally, the lower left panel shows ionization for medium
impact parameter $b=2.21$ a$_0$ and two different velocities (0.1
a.u. and 0.2 a.u.). Ionization increases with velocity although the
change is surprisingly small. We have here two counteracting
influences. The faster atom exposes the molecule higher frequency components so it can enhance ionization. On the other
hand, the time of interaction becomes much shorter leading to a
reduced reaction yield. The combination of these effects leads to a rather small velocity dependence in this range.

\subsubsection{Proton impact upon H$_2$O at low velocity}
\label{sec:res_theo_charged}
\begin{figure*}[htbp]
\begin{center}
\includegraphics[width=0.85\linewidth]{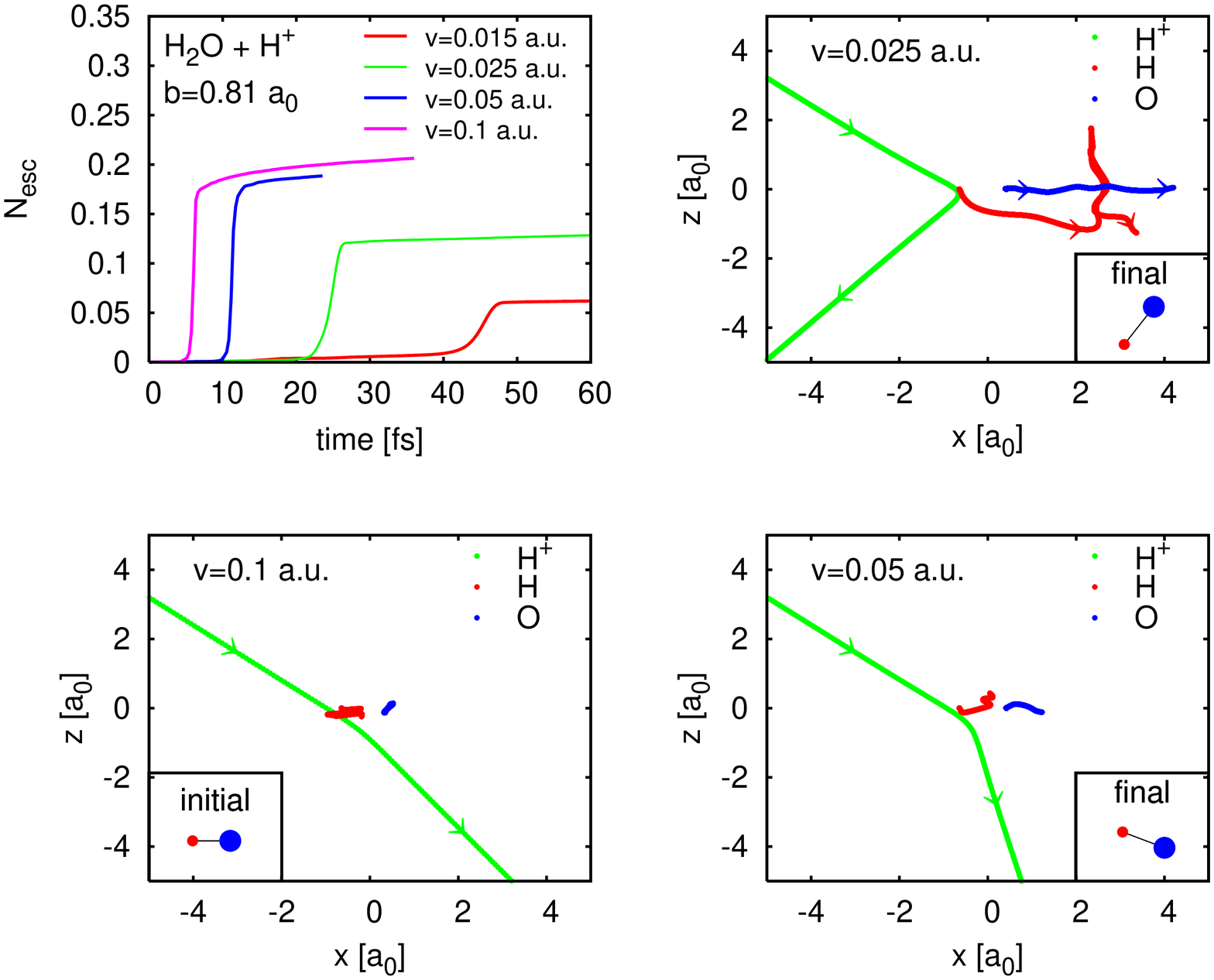}
\caption{Modeled collision between a water molecule and a H$^+$ ion at low
  velocities (v=0.015-0.1 a.u.) for a fixed impact parameter $b=0.81$
  a$_0$. Top left panel: number of escaped electrons as a function
  of time. Other panels: ionic trajectories in the $x$-$z$ plane. The
  arrows indicate the time evolution; only in the top right panel, the
  arrows on the water molecule are drawn at same instant.
  The insert in the bottom left panel shows the initial configuration
  of H$_2$O for all the calculations, while the other inserts present the
  final configuration for $v=0.05$ a.u. (bottom right) and $v=0.025$ a.u.
  (top right).}
\label{fig:lowH+}
\end{center}
\end{figure*}
The calculations treating charged projectile H$^+$
collisions with H$_2$O at a small impact parameter ($b=0.81$
a$_0$) and at various low velocities are shown in
Fig.~\ref{fig:lowH+}. The H$_2$O molecule is placed with
the O atom in the scattering plane and the two H atoms facing out
of the plane. The interaction between the charged projectile
and the molecule has a larger range than the neutral
projectile case due to ion charge coupling with the large dipole
moment of the water molecule. As a consequence, the
projectile trajectories experience bending depending on
the projectile velocity. The upper right panel shows the
lowest velocity case and it is clear that the bending angle is
very high. The water molecule as a whole is accelerated to a
translational motion combined with some rotation of the two H
atoms relative to the O atom (compare the initial and
final configurations shown in the inserts).
The lower right panel represents a collision at 0.05 a.u.. The
bending of the projectile trajectory and the effect on the motion
of the H$_2$O molecule are both markedly smaller than observed for
v=0.025 a.u.. Again, the molecule is accelerated to translational
motion with some small rotation, but this time in the
other direction.
The fastest collision is shown in the lower left panel. The much
shorter interaction time reduces bending and impact on
the molecule.
The left upper panel shows ionization for the various impact
velocities. The slowest collision, although having the largest effect
on molecular motion, produces the least electron emission.
Ionization increases with increasing velocity and seems to
level off near the largest velocity. Thus, the enhancement by
higher frequency components dominates over reduced interaction
time in that (low) velocity range.
 \\
 \\
The theoretical test cases have demonstrated the
feasibility of detailed dynamical simulations of water systems
excited by various projectiles. We have seen a
significant difference in the dynamics triggered by
neutral (C atom) versus  charged (H$^+$ ion) projectiles.
There is sizeable impact ionization. The effects on the molecular
motion as a whole are large and depend sensitively on the impact
conditions.

\section{Conclusion and outlook}

In this paper we have demonstrated the ability to produce intense beams of accelerated and size-selected water cluster ions in our newly constructed cluster/atom collision apparatus. The accompanying theoretical studies have demonstrated our ability to explore the
dynamics of water systems irradiated
by various projectiles. The next steps of the experimental investigation will focus on the effects of irradiation of size selected water cluster ions. The parallel theoretical investigations are currently being extended 
to consider irradiation of small singly charged water clusters. 
In particular, comparisons will be drawn between our experimental and theoretic studies  of size-selected water cluster ion collisions with high-velocity protons.
Attachment of water molecules to a molecule of biological 
interest will constitute a further step
both from the experimental and theoretical sides. The experimental capability to 
control the number of attached water molecules  is essential for a direct comparison with theoretical calculations. This opportunity will be  
exploited in the studies to come.

Acknowledgments. This work was supported by
the Deutsche Forschungsgemeinschaft (RE 322/10-1),
the  Humboldt foundation, a Gay-Lussac prize, Institut Universitaire de France,
Agence Nationale de la Recherche (ANR-06-BLAN-0319-02)

\end{document}